\documentclass[aps, prd, superscriptaddress, nofootinbib, preprintnumbers, twocolumn]{revtex4}
\usepackage{graphics}
\usepackage{epsfig,amsfonts,amssymb,amsmath}
\usepackage{epstopdf}
\usepackage{latexsym}
\usepackage{colordvi}
\usepackage{refstyle}
\usepackage{float}
\newcommand{\nn}{\nonumber}
\newcommand{\be}{\begin{equation}}
\newcommand{\ee}{\end{equation}}
\newcommand{\bea}{\begin{eqnarray}}
\newcommand{\eea}{\end{eqnarray}}
\begin{document}

\title{RG flow of AC Conductivity in Soft Wall Model of QCD}
\author{Neha Bhatnagar}
\email{bhtngr.neha@gmail.com}
\author{Sanjay Siwach}%
 \email{sksiwach@hotmail.com}
\affiliation{%
 Department of Physics,Banaras Hindu University,\\Varanasi-221005, India\\
}%

\begin{abstract}
We study the Renormalization Group (RG) flow of AC conductivity in soft wall model of holographic QCD. We consider the charged black hole metric and the explicit form of AC conductivity is obtained at the cut-off surface. We plot the numerical solution of conductivity flow as a function of radial co-ordinate. The equation of gauge field is also considered and the numerical solution is obtained for AC conductivity as a function of frequency. The results for AC conductivity are also obtained for different values of chemical potential and Gauss-Bonnet couplings.\\

{\bf Keywords:}  Holographic QCD; Soft wall model; Transport Properties; Gauss-Bonnet coupling
\end{abstract}


\maketitle

\section{\label{sec:level1}Introduction}

AdS/CFT correspondence\cite {Maldacena:1997re} is widely used to study the dynamics of strongly correlated systems. AdS/CFT correspondence dictates that the dynamics of boundary theory can be studied by considering the fields in the dual gravity theory in the bulk AdS spacetime. A serious application of these ideas in QCD started from the work of Sakai and Sugimoto \cite{Sakai:2004cn}. 

Alternatively, the phenomenological models of QCD \cite{Erlich:2005qh}, makes use of some known features of QCD and tried to incorporate these features in the bulk gravity theory. In these model the phenomena like chiral symmetry breaking are introduced by bifundamental fields dual to chiral condensate and confinement is realized by introducing an IR cut-off. This is further modified to include the Regge trajectories of mesons in the model and popularly known as soft wall model of QCD \cite{Karch:2006pv}. The role of IR cut-off is played by a dynamical wall in this model.

These models are studied further to investigate the phase structure and thermodynamics of QCD \cite{Sachan:2011iy}. The transport properties like AC and DC conductivity and diffusion constant are also investigated in these models. \cite{Son:2002sd,Kovtun:2003wp,Erlich:2009me,Jain:2009bi,Edalati:2010hk,Kim:2010ag, Albrecht:2012ek,Donos:2012js}. Renormalization Group (RG) flow of transport co-efficients in theories dual to charged black hole is studied using membrane paradigm \cite{Iqbal:2008by,Faulkner:2010jy,Heemskerk:2010hk,Bredberg:2010ky,Sin:2011yh,Matsuo:2009yu,Matsuo:2011fk,Lee:2011qu,Ge:2014eba},where the dynamics of vector and tensor perturbations is used to study the flow equations for conductivity and diffusion constant. Here we adapt this formalism \cite{Iqbal:2008by,Matsuo:2011fk,Ge:2014eba} to investigate the re-normalization flow of AC conductivity in the soft wall model of QCD. This is further generalized to include the effects of Gauss-Bonnet couplings in the bulk.

The paper is organized as follows. First, we consider the charged black hole solution of Einstein-Maxwell theory and perturbation equations for the bulk metric and gauge field in the soft wall model. We calculate the AC conductivity using the membrane paradigm and explicit results are given in the near horizon limit. We also plot the full solution for AC conductivity as a function of frequency and the results in the probe limits are also plotted. Secondly, the results are obtained for AC  conductivity while considering higher order gravity corrections in the action known as Gauss-Bonnet corrections. We also give explicit expression for DC conductivity at the cut off surface in the appendix. 

\section{Transport Coefficients in Einstein-Maxwell Theory}

Let us consider the Einstein-Maxwell action in 5-dimensions,
\begin{equation}\label{eq:action0}
S=\int~d^5x~\sqrt{-g}e^{-\phi}~\left\lbrace\frac{1}{2\kappa^2}\left(R-2\Lambda\right)-\frac{1}{4g^2}F^2\right\rbrace,
\end{equation}
where $F^2=F_{mn}F^{mn}$ is the Lagrangian density of the Maxwell field, and $\phi$ is the dilaton field. The constant $\kappa^2$ is related to five dimensional Newton's constant $G_5$ as $\kappa^2=8\pi G_5$ , and the cosmological constant is related to AdS radius, $\Lambda=-6/l^2$. The AdS/QCD correspondence relates five dimensional gravitational constant $2\kappa^2$ and five dimensional gauge coupling to the rank of color group ($N_c$) and number of flavours ($N_f$) in the boundary theory,
$\frac{1}{2\kappa^2}=\frac{N_c^2}{8\pi^2}$,~~~ $\frac{1}{2g^2}=\frac{N_cN_f}{4\pi^2}$.

The charged black hole solution\cite{Chamblin:1999tk,Ge:2008ak,Matsuo:2011fk} of Einstein-Maxwell gravity in five dimensions with negative cosmological constant is given by,
\bea
ds^2&=&\frac{r^2}{l^2}(-f(r)dt^2+\sum_{i=1}^3 dx^idx^i)+\frac{l^2}{r^2f(r)}dr^2, \\
A_t&=&\mu(1-\frac{r_+^2}{r^2}) \nn
\eea
where
$f(r)=1+a\frac{r_+^6}{r^6}-(1+a)\frac{r_+^4}{r^4}$, and the charge of the black hole is related with parameter `a' and chemical potential $\mu$ as:   $a=\frac{l^2\kappa^2Q^2 }{6g^2}$ , $Q=\frac{2\mu}{r_+}$. \\ ~~~\\

Defining,  $u=\frac{r^2_+}{r^2}$ for simplification, the above metric can be written as,
\be
ds^2=\dfrac{r_{+}^2}{l^2u}(-f(u)dt^2+\sum_{i=1}^3 dx^idx^i)+\dfrac{l^2du^2}{4u^2f}
\ee
where $f(u)=(1-u)(1+u-au^2)$. 

We take dilaton field in soft wall model as $\phi=cu$ and the constant $c=0.388 $GeV$^2$ \cite{Karch:2006pv}.

The equation of motion for the gauge field in the soft wall models can be written as, 
\be\label{eq:gauge}
\frac{1}{\sqrt{-g}}\partial_m(\sqrt{-g}e^{-\phi}\, F^{mn})~=~0. \\
\ee
Let us consider the metric and gauge field perturbations as,
\bea
 g_{mn}~=~g^0_{mn}+\tilde{h}_{mn}\\
\textit{A}_{m}~=~A_{m}^0+A_{m}.
\eea
We scale the metric perturbation as $\tilde{h}_{mn}=e^{\phi}{h}_{mn}$
and take the Fourier decomposition of the fields as follows,
\bea \label{eq:eqft}
h_{mn}(t,z,u)~=~\int\frac{d^4k}{(2\pi)^4}e^{-i\omega t+ikz}h_{mn}(k,u)\\
A_m(t,z,u)~=~\int\frac{d^4k}{(2\pi)^4}e^{-i\omega t+ikz}A_m(k,u).
\eea
Now, focusing on the linearised theory for $h_{mn}$ and for vector field $A_m$ propagating in charged black hole background with the gauge condition $h_{un}=0$ and $A_u=0$, one gets the equation of motion for vector modes of metric perturbations  $ h^x_z$\textrm{and} $h^x_t $   
\bea
0=kfh'^{x}_z+\omega h'^{x}_t-3a\omega uA_x \label{eq:perti} \\
0=h''^{x}_t-\frac{1}{u}h'^{x}_t-\frac{b^2}{uf}(\omega  kh^x_z+k^2h^x_t)-3auA'_x\label{eq:pert1} \\
0=h''^{x}_z+\frac{(u^{-1}f)'}{u^{-1}f}h'^x_z+\frac{b^2}{uf^2}(\omega^2h^x_z+\omega kh'^{x}_t) \label{eq:pertii} 
\eea
where, $b=\frac{l^2}{2r_+}$. 
Similarly, the gauge field \eqref{gauge}, becomes,  
\be
0=A''_x+(\frac{f'}{f}-c) A'_x+\frac{b^2}{uf^2}(\omega^2-k^2f)A_x-e^{\phi}\frac{A'_t}{f}h'^{x}_t \label{eq:cd}.
\ee
The gauge field equation is coupled with metric perturbations in charged black hole background and one has to resort to numerical methods to solve these equations in order to calculate the AC conductivity. However, in the near horizon regime an exact solution can be obtained.  

We consider AC conductivity flow in the soft wall model in membrane paradigm \cite{Iqbal:2008by} and using the fact that,
\be\label{eq:cond}
\sigma_{A}(\omega,u)~=~\dfrac{J^x}{i\omega A_{x}},
\ee
where the current density is given as,
\be\label{eq:gauge1}
 J^{x}~=~\dfrac{-1}{g^2}\sqrt{-g}e^{-\phi}F^{ux}+\frac{g_{xx}}{2\kappa^2}\sqrt{-g}A'_th^x_t.
\ee
Now, using \eqref{gauge,cond,gauge1} the conductivity flow can be written as, 
\bea
\dfrac{\partial_{u_{c}}\sigma_{A}}{i\omega}-\dfrac{g^2\sigma_{A}^2}{\sqrt{-g}e^{-\phi}g^{uu}g^{xx}}-\dfrac{2\kappa^2g^{xx}g^4\omega^2}{\sqrt{-g}}\dfrac{4 u^3(A'_t)^2 }{r_+^2}\nn \\
+\dfrac{\sqrt{-g}e^{-\phi}g^{xx}g^{tt}}{g^2}~=~0.
\eea
\begin{figure}[H]
\centering
\includegraphics[width=0.45\textwidth]{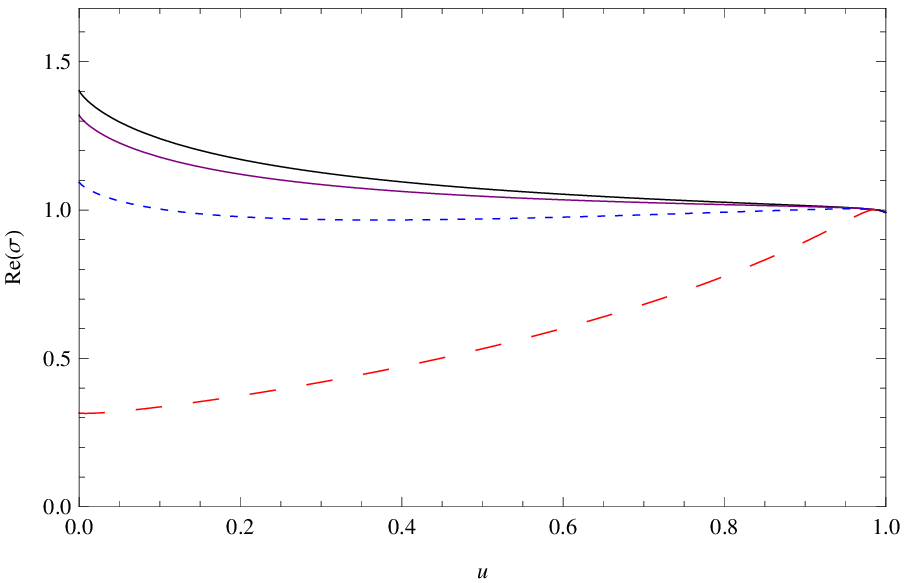} 
\includegraphics[width=0.45\textwidth]{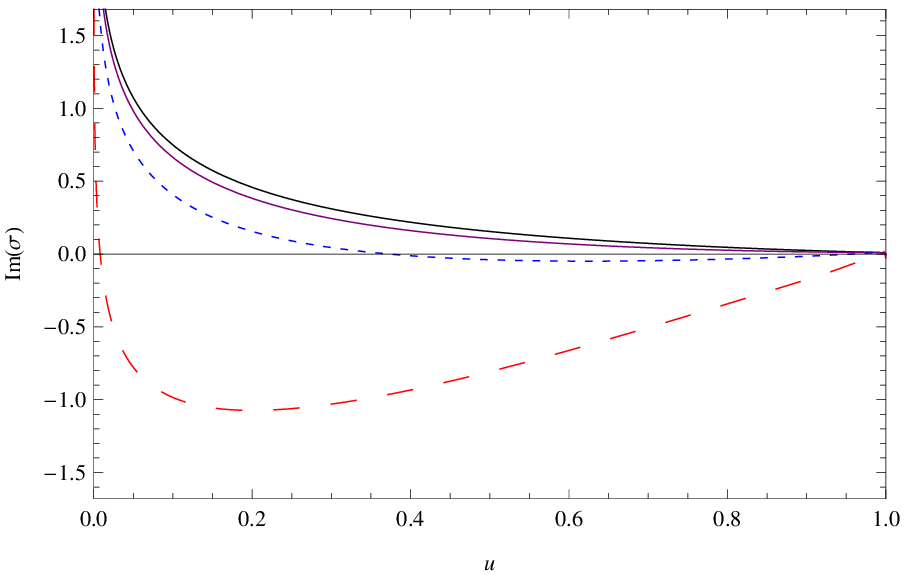}
\caption{The radial flow of AC conductivity($\sigma$) at fixed frequency($\omega=0.999$) with varying chemical potential ($\mu$=0.01(black), 0.05(magenta), 0.1(dotted- blue), 0.25(dashed red))}
 \label{fig:fig1} 
\end{figure}
In the near horizon limit, $\sigma_A$ is a constant and can be evaluated by applying regularity condition at the horizon, (u=1)
\be
\sigma_A(u=1)~=~\dfrac{e^{-c}}{g^2}\dfrac{r_+}{l}.
\ee
RG flow plots for AC conductivity has been shown in Fig.1. The frequency dependence of AC conductivity has been evaluated numerically and shown in Fig.3, Fig.4, Fig.5 and Fig.6. In the probe brane limit, taking $f(u)=1-u^2$, the plots in Fig. 3, show striking similarity with condensed matter systems \cite{Hartnoll:2008vx}.

\section{Transport Coefficients with Gauss Bonnet Corrections}

We study the effect of Gauss-Bonnet(GB) coupling on the RG flow of conductivity in the soft wall model approach. The modified action with the GB term is given by,

\be\label{eq:action1}
S~=~\int~d^5x~\sqrt{-g}e^{-\phi}~\left\lbrace\frac{1}{2\kappa^2}\left(R-2\Lambda+\alpha
R_{GB}\right)-\frac{1}{4g^2}F^2\right\rbrace,
\ee

where $R_{GB}~=~R^2-4R_{MN}R^{MN}+R^{MNPQ}R_{MNPQ}$ is Gauss-Bonnet term
and `$\alpha$' is the Gauss Bonnet Coupling constant.  

We consider the solution for Einstein-Maxwell-Gauss-Bonnet(EMGB) system as \cite{Cai:2001dz,Cai:2009zv,Buchel:2009sk,Ge:2011fb,Hu:2011ze}.

\be
ds^2~=~\dfrac{r_{+}^2}{l^2u}(-f(u)N^2dt^2+\sum_{i=1}^3 dx^idx^i)+\dfrac{l^2du^2}{4u^2f},
\ee
where
\bea
N^2&=&\frac{1}{2} \left(\sqrt{1-4 \alpha }+1\right) \nn \\
f(u)&=&\dfrac{1}{2\lambda}(1-\sqrt{1-4\lambda(1-u)(1+u-au^2)}, \nn
\eea

and $\lambda$ is related with the Gauss Bonnet coupling term as $\lambda~=~\frac{\alpha}{l^2}$.

Using the membrane paradigm as explained above for the charged black hole case and perturbation of metric and gauge fields as in previous section, the modified equations of motion for vector perturbations are given by,  

\be
0= \omega h'^{x}_t-\frac{uM'}{M}fN^2kh'^x_z\\-\frac{3aN^2A_x\omega}{M}
\ee
\be
0= h''^{x}_t+\frac{M'}{M}h'^{x}_t+\frac{l^4b^2}{4f}\frac{M'}{M}(\omega kh^x_z+k^2h^{x}_t)-\frac{3aN^2A'_x}{M}
\ee
\be
\begin{split}
0= h''^{x}_z-\frac{\frac{1}{u}-\frac{f'}{f}+2\lambda[\frac{f}{u}+uf''+u\frac{f'^2}{f}-2f']}{1+2\lambda(uf'-f)}h'^x_z \\+ \frac{l^4b^2}{4N^2uf^2}(\omega^2h^x_z+\omega kh^{x}_t),
\end{split}
\ee
\be\label{eq:gauge3}
0= A''_x+(\frac{f'}{f}-c)A'_x+\frac{l^4}{4N^2f^2r_+^2u}(\omega^2-k^2f)A_x-e^\phi\frac{1}{N^2f}h'^{x}_t 
\ee

where $M~=~\frac{1-2\lambda f(u)}{u}$.  
  

In order to determine the flow equation of AC conductivity, we consider the current density with the GB corrections defined as,
\be
J^x=-\frac{1}{g^2}\sqrt{-g}g^{uu}g^{xx}e^{-\phi}\partial_uA_x +\frac{1}{g^2}\sqrt{-g}\frac{4u^3}{r_+^2 N^2}A'_th^x_t 
\ee
The corresponding RG flow equation for AC conductivity (using definition \eqref{cond})becomes,

\be\label{eq:gauge2}
\begin{split}
\dfrac{\partial_{u_{c}}\sigma_{A}}{i\omega}-\dfrac{g^2\sigma_{A}^2}{\sqrt{-g}e^{-\phi}g^{uu}g^{xx}}-\dfrac{2\kappa^2 g^{xx}\sqrt{-g}}{g^4\omega^2}(A'_t)^2 \dfrac{4 u^2}{Mr_+^2} \\
+\dfrac{\sqrt{-g}e^{-\phi}g^{xx}g^{tt}}{g^2}=0 
\end{split}
\ee

In the near horizon limit, $u_c~=~1$(the cut-off horizon), we can get an exact expression for the AC conductivity,
\bea
\sigma_A(u_c=1)~=~\dfrac{r_+}{g^2}\dfrac{e^{-c}}{l}
\eea
RG flow plots for AC conductivity with GB corrections has been shown in Fig.3 and we can notice that the qualitative feature of the flow are similar to the case without Gauss Bonnet correction. \\ ~~~\\

\begin{figure}[H]
\centering
\includegraphics[width=0.45\textwidth]{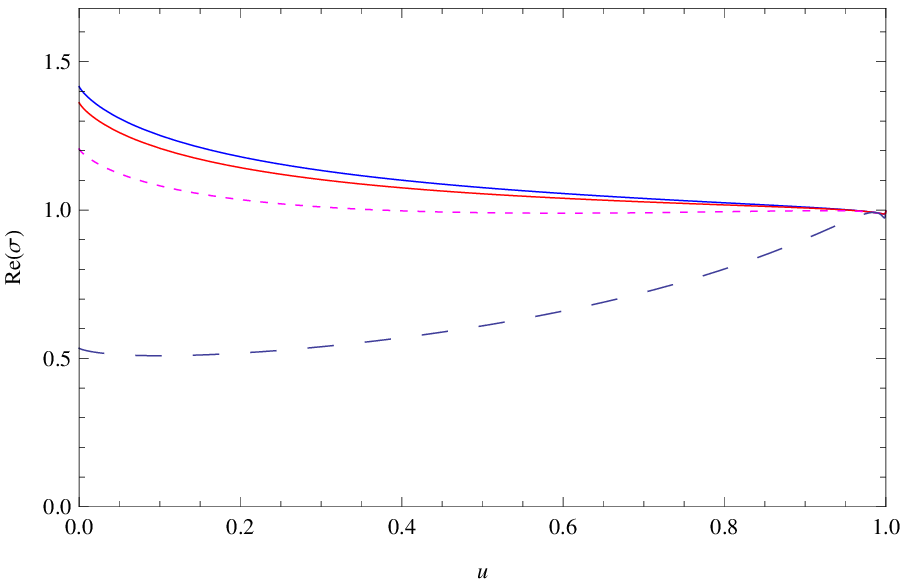} 
\includegraphics[width=0.45\textwidth]{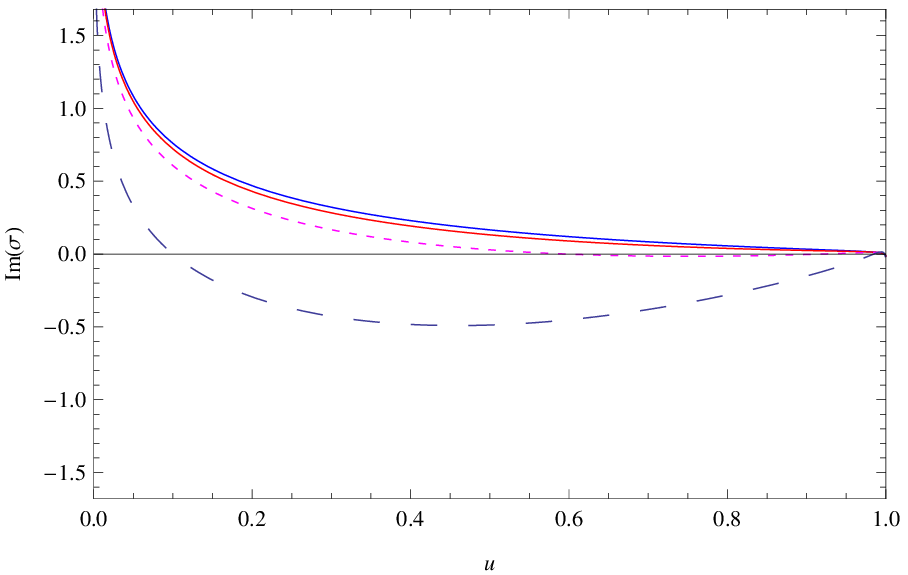} 
\caption{The radial flow of AC conductivity($\sigma$) with GB corrections($\lambda =0.01$) at fixed frequency($\omega$=0.999) with differnet $\mu$=0.01(blue), 0.05(Red), 0.1(dashed magenta), 0.25(dashed blue)}
\label{fig:fig3} 
\end{figure} 

\begin{figure*}
\includegraphics[width=0.45\textwidth]{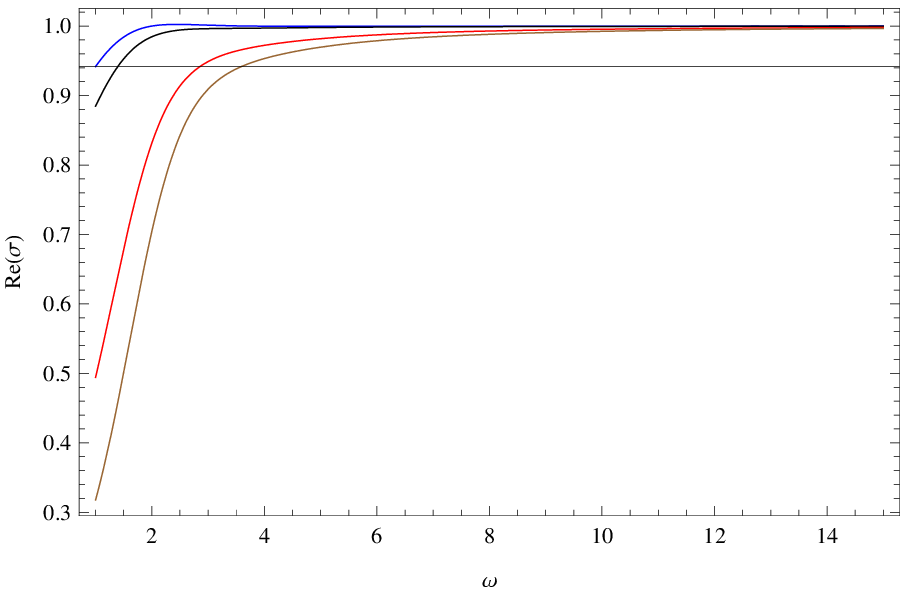}
\includegraphics[width=0.45\textwidth]{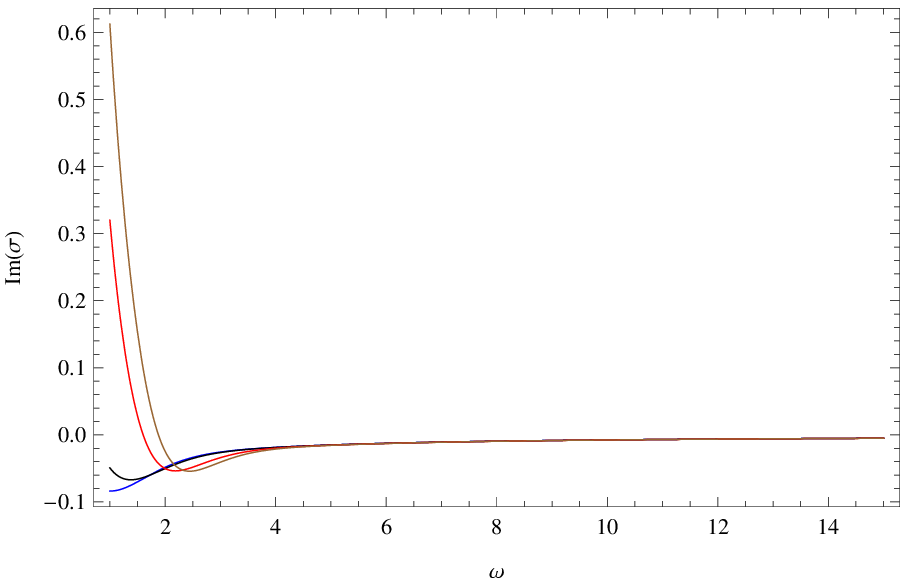}  
\caption{Frequency dependence of AC conductivity($\sigma$) in the probe limit  with varying chemical potential ($\mu=$ 0.01(blue), 0.1(black), 0.9(red), 1.5(brown))}
\label{fig:fig8}
\end{figure*}
 
\section{Conclusions}

The soft wall model of Holographic QCD is used here to get the insights into the transport properties like AC and DC conductivity in strongly coupled regime of QCD. The flow equations of AC and DC conductivity are considered for different values of chemical potential. The numerical solution of these equations enabled us to calculate the value of real and imaginary part of AC conductivity and the results seem to agree with the models, which consider the dynamics of condensate. This suggests that the soft wall model successfully captures the same features. This has also been noticed recently by \cite{Afonin:2015fga} independently. In the probe limit our results (Fig.3) agree with existing results in the literature \cite{Hartnoll:2008vx,Hartnoll:2009sz,Matsuo:2011fk}. The results at high frequency (Fig.4, Fig.5) show oscillatory behavior of AC conductivity, which is reminiscent of Shubnikov de Haas effect. At low frequency, the Drude behavior\cite{Donos:2012js} is observed (Fig.6). The Gauss-Bonnet coupling does not change the results significantly. \\

\section{Acknowledgement}
We acknowledge the financial support from the DST, Govt. of India, Young Scientist project. 

\appendix*

\section{DC CONDUCTIVITY}

In order to calculate DC conductivity, we consider the gauge field equation \eqref{cd} in the limit $\omega,k =0$ the equation of motion for gauge-field becomes,
\be
A_x''+(\frac{f'}{f}-c)A_x'-e^{\phi}\frac{A'_t}{f}h'^{x}_t~=~0
\ee
The solution of $A_x(u)$, can be written as,
\be
A_x(u)~=~A_x(0)(1+\dfrac{3 a u (e^{c u}-1)}{cf'-3 a u(e^{c u}-1)}),
\ee
which can be used to calculate DC conductivity using the expression given in  \cite{Iqbal:2008by, Matsuo:2011fk}
\be
\sigma_{DC}=\dfrac{\sqrt{-g}}{g^2}g^{xx}g^{uu}e^{-\phi}\sqrt{\dfrac{g_{uu}}{g_{tt}}}\vert_{u=1}\dfrac{A_x(1)A_x(1) }{A_x(u_c) A_x(u_c)}.
\ee
Thus, we get the following expression for DC conductivity flow,
\bea
\sigma_{DC} = e^{-c}\frac{r_+}{g^2 l}\frac{(a-2)^2 \left(a \left(-3 c u_c+3 e^{c u_c}+2 c-3\right)+2 c\right){}^2}{\left(a \left(c-3 e^c+3\right)-2 c\right)^2 \left(a \left(3 u_c-2\right)-2\right){}^2}\nn
\eea
In the near horizon region, $u_c=1$, the DC conductivity takes the form,
\be
\sigma_{DC}~=~\dfrac{r_+e^{-c}}{g^2l}.
\ee
At the boundary $u_c=0$, DC flow is given as
\be
\sigma_{DC}~=~\frac{r_+}{g^2 l}e^{-c}\frac{(a-2)^2 c^2}{\left(a \left(c-3 e^c+3\right)-2 c\right)^2}.
\ee

Following the above procedure, we can consider the DC conductivity with Gauss-Bonnet corrections. We obtain the identical expression for DC conductivity indicating that the flow is independent of Gauss-Bonnet terms.\\

\begin{figure*}
\includegraphics[width=0.45\textwidth]{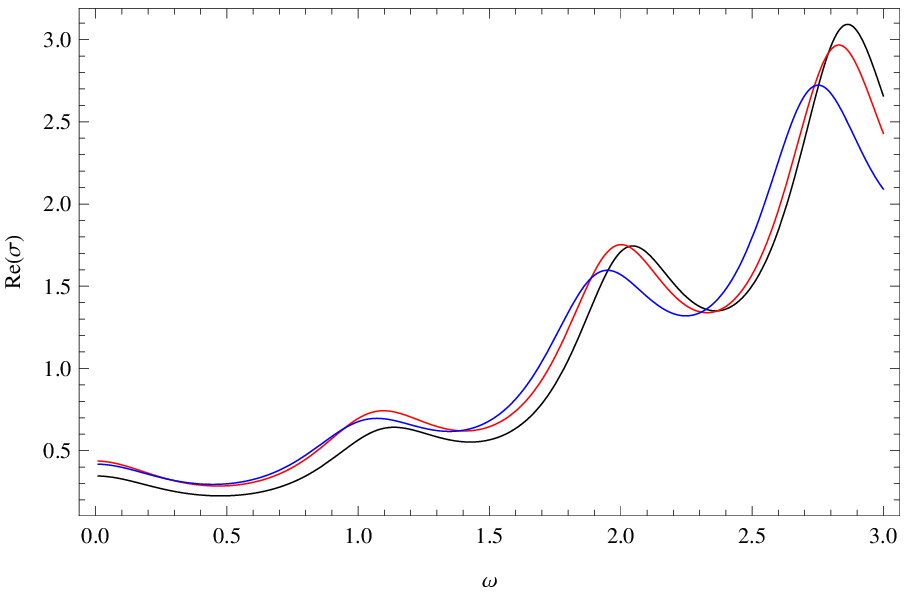} 
\includegraphics[width=0.45\textwidth]{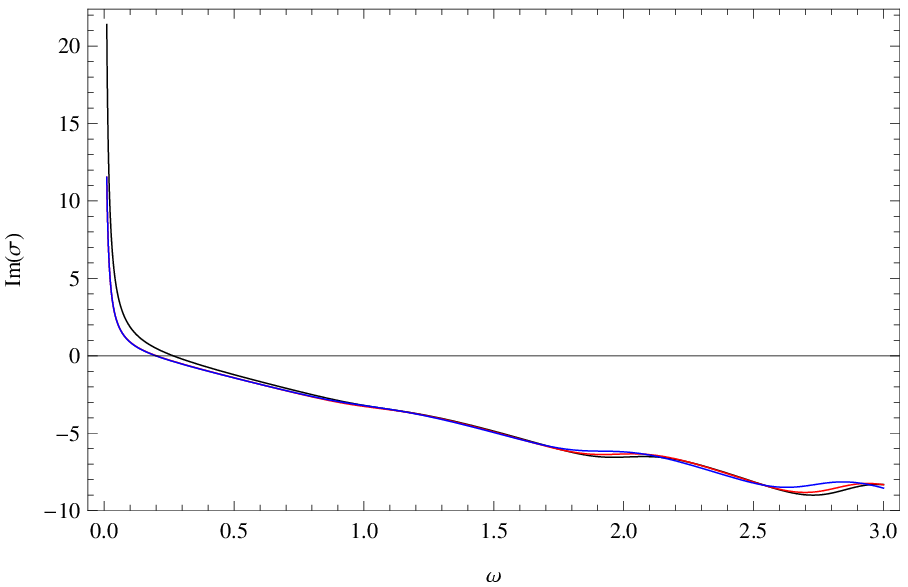} 
\caption{Frequency dependence of  AC conductivity ($\sigma$) for fixed chemical potential ($\mu$=0.01) and varying Gauss-Bonnet terms ($\lambda$ =0(black), 0.01(red), 0.05(blue))}
\label{fig:fig5}
\end{figure*}

\begin{figure*}
\includegraphics[width=0.45\textwidth]{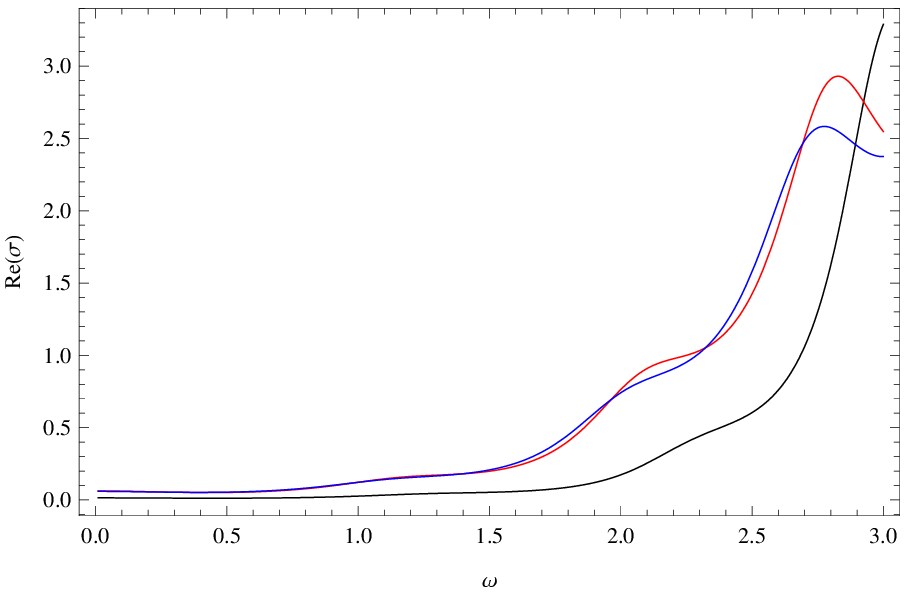} 
\includegraphics[width=0.45\textwidth]{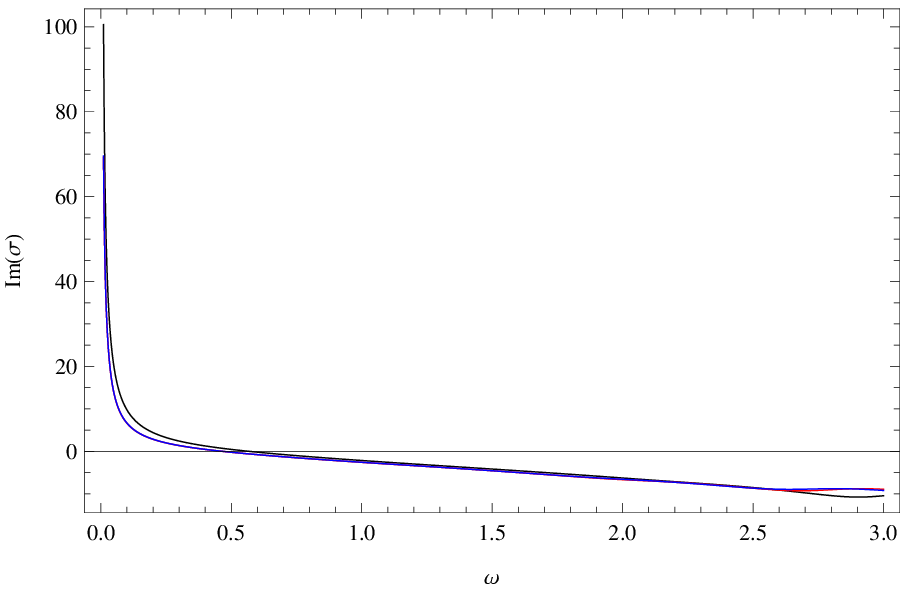} 
\caption{Frequency dependence of AC conductivity ($\sigma$) for higher chemical potential ($\mu$=0.1) and varying Gauss-Bonnet term ($\lambda$ = 0(black), 0.01(red), 0.05(blue)) }
\label{fig:fig6}
\end{figure*}

\begin{figure*}
\centering
\includegraphics[width=0.45\textwidth]{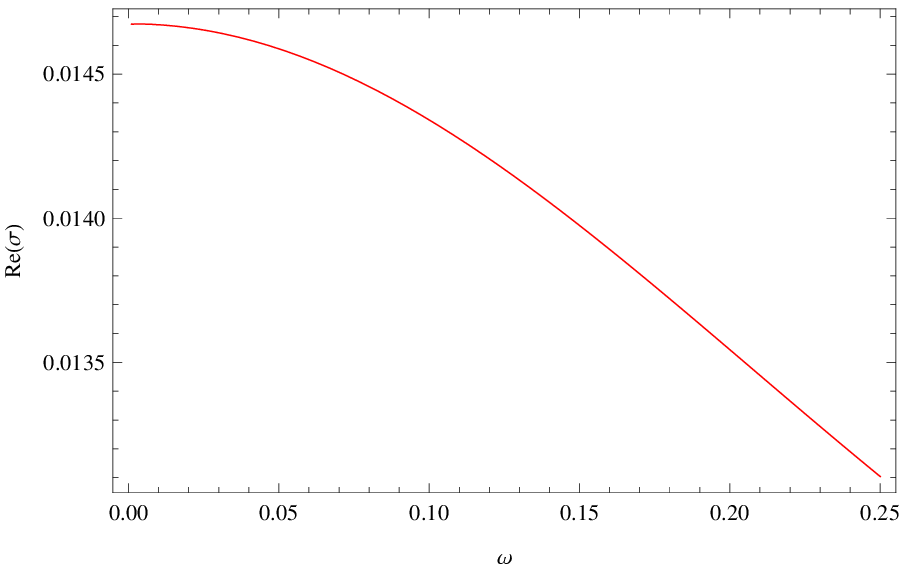} 
\includegraphics[width=0.45\textwidth]{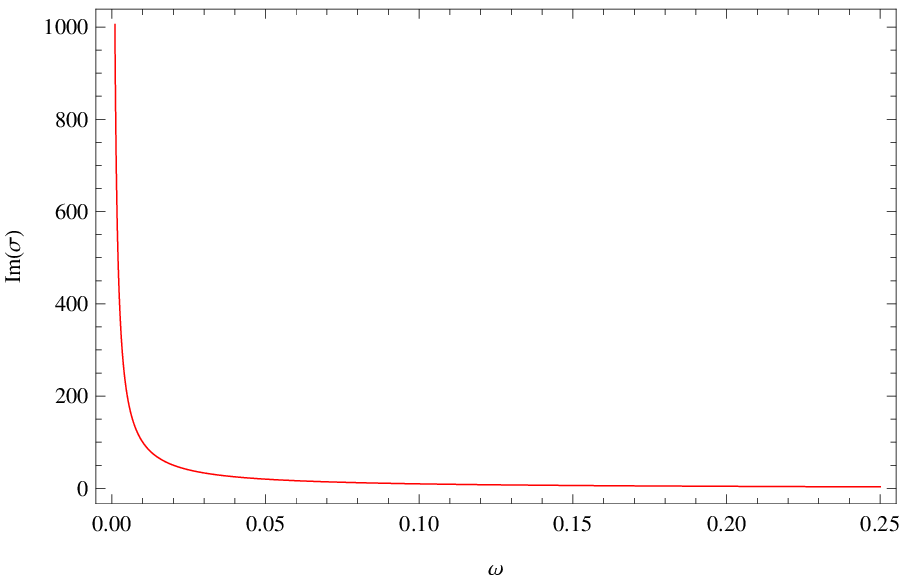}
\caption{Frequency(low) dependence of AC conductivity($\sigma$) for fixed chemical potential ($\mu$=0.1)}
\label{fig:fig7}
\end{figure*}

\end{document}